\documentclass[aps,amssymb,prd,showpacs,twocolumn]{revtex4}
\usepackage{amsmath,bm}
\begin{document}

\title{Potential for ill-posedness in several 2nd-order formulations of the 
Einstein equations}

\author{Simonetta Frittelli}
\email{simo@mayu.physics.duq.edu}
\affiliation{Department of Physics, Duquesne University,
       Pittsburgh, PA 15282}
\affiliation{Department of Physics and Astronomy, University of Pittsburgh,
       Pittsburgh, PA 15260}

\date{\today}

\begin{abstract}

Second-order formulations of the 3+1 Einstein equations obtained by 
eliminating the extrinsic curvature in terms of the time derivative of the
metric are examined with the aim of establishing whether they are well posed,
in cases of somewhat wide interest, such as ADM, BSSN and generalized
Einstein-Christoffel.  The criterion for well-posedness of second-order systems
employed is due to Kreiss and Ortiz. By this criterion, none of the three cases
are strongly hyperbolic, but some of them are weakly hyperbolic, which means
that they may yet be well posed but only under very restrictive conditions for
the terms of order lower than second in the equations (which are not studied
here).  As a result, intuitive transferences of the property of well-posedness
from first-order reductions of the Einstein equations to their originating
second-order versions are unwarranted if not false. 

\end{abstract}
\maketitle

\section{Introduction}\label{sec:1}

A common practice to study the well-posedness of a time-dependent second-order
system of partial differential equations is to reduce the system to first
order. This is done because first-order systems of PDE's have been amply
examined and are currently very well understood. By reducing to first-order,
what is meant is to find a first-order system of PDE's whose solution space
contains a subset that is equivalent to the solutions of the original
second-order system by a trivial identification.  If the first-order reduction
is well posed, then the solutions of the first-order system are bounded in
terms of the initial data of the first-order system, and theorems of existence,
uniqueness and stability under small perturbations then follow.   

It is tempting to presume that the well-posedness of a first-order reduction is
a sign of well-posedness in the originating second-order system, especially
because the originating second-order system is much smaller (in terms of the
number of variables) and may be more amenable to a numerical implementation.
However, in general relativity, the existence of constraints leads to
first-order reductions that are equivalent to the original second order problem
with regards to \textit{constrained solutions\/} only.  The evolution equations
of two first-order reductions of the Einstein equations that differ by linear
combinations with the hamiltonian or momentum constraints are not equivalent.
Thus, whether a well-posed first-order reduction of the Einstein equations
guarantees the well-posedness of the corresponding second-order problem is not
at all clear.  

The quintessential model of a well-posed time-dependent second-order equation
in three dimensions  is the wave equation in flat space. By the well-posedness
of the wave equation, more complicated systems of PDE's can be shown to be well
posed if they consist of series of wave equations, as is the case of the 3+1
Einstein equations in the harmonic gauge or time-harmonic
gauge~\cite{choquet83}.  But beyond the wave equation, little is known about
general well-posedness criteria for  systems of second-order time dependent
PDE's. 

A criterion for the well-posedness of time-dependent second-order systems of
PDE's has been developed recently by Kreiss and Ortiz~\cite{ortiz}. Even though
the criterion applies to cases that are much more general than what we are
interested in, for our purposes it can be stated as follows.  Consider a system
of $n$ second-order partial differential equations with constant coefficients. 
The system is of the form
\begin{equation}\label{1}
 	\ddot{\bm{u}}  = \sum_{jk}\bm{A}_{jk} \bm{u},_{jk}
\end{equation}

\noindent where $\bm{u}$ is a vector containing the $n$ fundamental variables 
and, for each value of $(j,k)$, $\bm{A}_{jk}$ is an $n\times n$ constant
matrix. An overdot denotes a partial derivative with respect to the time
coordinate ($\partial/\partial t$), so that $\ddot{f}\equiv \partial^2
f/\partial t^2$,  and $,_j\equiv \partial/\partial x^j$. For any arbitrary unit
covector $\xi_i \equiv \omega_i/|\omega|$, we define the $n$-dimensional matrix
$\bm{P}_0(\xi) \equiv \sum_{jk}\bm{A}_{jk}\xi_j\xi_k$. The system is strongly
hyperbolic if and only if the eigenvalues of  $\bm{P}_0(\xi)$ are strictly
positive and $\bm{P}_0(\xi)$ has a complete set of eigenvectors which is
uniformly (in $\xi$) linearly independent. 

This criterion is entirely similar to strong hyperbolicity of first-order
systems~\cite{kreissbook} except for the condition that the eigenvalues must be
\textit{strictly positive,\/} which excludes the value zero as well as all
negative values.  The reason for positivity of the eigenvalues is, essentially,
that each eigenvalue must allow for two waves travelling with the same speed
but in opposite directions (so, in a sense, the eigenvalues are the squares of
real characteristic speeds of any sign).  The reason for excluding the
vanishing eigenvalue is that there is only one associated ``wave'' with zero
speed instead of two.  This criterion emulates fully the case of the
three-dimensional wave equation.  Any second-order system of PDE's that
satisfies this criterion is well posed in the standard sense, that is: its
solutions are bounded by the initial data irrespective of the spectral
frequency of the data~\cite{ortiz}. On the other hand, if the eigenvalues are real and
non-negative and/or there isn't a complete set of eigenvectors, the system is
referred to as \textit{weakly hyperbolic}.  According to \cite{ortiz}, weakly
hyperbolic systems can develop ``catastrophic exponential growth'' when adding
lower order terms or considering variable coefficients. It is inferred that if
the linearization of a system of nonlinear second-order equations around a
constant background is weakly hyperbolic, then the associated non-linear system
itself is prone to ``catastrophic exponential growth'', as is its linearization
around any background that is not constant. Examples of what Kreiss and Ortiz
refer to as ``catastrophic growth'' appear in \cite{ortiz}. 

As far as we are aware of, the criterion has been developed only for linear
systems with constant coefficients, but there is reason to presume that it can
be generalized to variable coefficients and quasilinear systems in a manner
similar to the case of first-order systems.  Therefore, in the following, we
will use this criterion to analyze the well-posedness of second-order
formulations of the Einstein equations insofar as they are linearized around
flat space. The case of the standard ADM equations~\cite{yorksources} is
studied in Section~\ref{sec:2}. The case of the widely used BSSN
equations~\cite{BSSN} is dealt with in Section~\ref{sec:3}. Finally, the case
of the generalizeded Einstein-Christoffel equations (EC)~\cite{KST}, which
includes the case of the Einstein-Christoffel equations
themselves~\cite{fixing}, is developed in Section~\ref{sec:4}.  We find that
none of the second-order versions of the equations is strongly hyperbolic. The
relevance of this result is summarized in Section~\ref{sec:5}.

\section{The ADM equations in second order form\label{sec:2}}

Throughout the article we assume the following form for the metric of spacetime
$g_{ab}$ in coordinates $x^a = (x^i,t)$ in terms of the three-metric
$\gamma_{ij}$ of the slices at fixed value of $t$:
\begin{equation}\label{metric}
    ds^2 = -\alpha^2 dt^2 + \gamma_{ij}dx^idx^j \;,
\end{equation}

\noindent where $\alpha$ is the lapse function. The Einstein equations
$G_{ab}=0$ for the four-dimensional metric are equivalently expressed in the
ADM form~\cite{yorksources}:
\begin{subequations}\label{adm}
\begin{eqnarray}
    \dot{\gamma}_{ij} &=& - 2\alpha K_{ij}, \label{adma}\\
    \dot{K}_{ij} &=& \alpha \left(R_{ij} - 2 K_{il}K^l{}_j
            +K K_{ij}\right) -D_iD_j\alpha, \label{admb}
\end{eqnarray}
\end{subequations}

\noindent with the constraints
\begin{subequations}\label{admconst}
\begin{eqnarray}
    {\cal C} &\equiv &\frac12\left(R - K_{ij}K^{ij} + K^2 \right)= 0,   \\
    {\cal C}_i&\equiv &D_jK^j{}_i-D_i K = 0,
\end{eqnarray}
\end{subequations}

\noindent to be imposed on the initial data. Indices are raised with the
inverse metric $\gamma^{ij}$, $D_i$ is the covariant three-derivative
consistent with $\gamma_{ij}$ , $R_{ij}$ is the Ricci curvature tensor of
$\gamma_{ij}$, $R$ its Ricci scalar, $K_{ij}$ is the extrinsic curvature of the
slice at fixed value of $t$ and $K\equiv \gamma^{ij}K_{ij}$. This system of
evolution equations is a (partial) first-order reduction of the original
second-order Einstein equations, which we can recover by substituting in
(\ref{admb}) the extrinsic curvature in terms of the time derivative of the
metric as given by (\ref{adma}):
\begin{equation}
    \ddot{\gamma}_{ij} = -2\alpha^2 R_{ij}
+\gamma^{lm}\dot{\gamma}_{il}\dot{\gamma}_{ml}
-\frac12 \gamma^{lm}\dot{\gamma}_{lm}\dot{\gamma}_{ij}
+\frac{\dot{\alpha}}{\alpha}\dot{\gamma}_{ij} +2\alpha D_iD_j\alpha. 
\label{admsecond}
\end{equation}

\noindent Linearizing around flat space, so that $\gamma_{ij} =
\delta_{ij}+h_{ij}$ and $\alpha=1+\epsilon$, these equations read
\begin{equation}\label{admlin}
    \ddot{h}_{ij} = \delta^{kl}
		   (h_{kl,ij}-h_{il,kj}-h_{jl,ki}+h_{ij,kl})
+2 \epsilon,_{ij}. 
\end{equation} 

\noindent Obviously the problem depends on the specification of the lapse
function.  We are here interested in two special cases: the case of 
constant unit lapse and the case of lapse equal to the square root of the
determinant of the three-metric. 

In the first place, then, suppose $\epsilon = 0$. The equations then read
\begin{equation}
    \ddot{h}_{ij} = \delta^{kl}
		   (h_{kl,ij}-h_{il,kj}-h_{jl,ki}+h_{ij,kl})\label{admlingeo}
\end{equation} 

\noindent and conform to Eq.~(\ref{1}). Solving the eigenvalue problem of
the principal symbol $P_0(\xi)$ is equivalent to making the ansatz $h_{ij} =
V_{ij} \exp i(\xi_kx^k -st)$ for an arbitrary covector $\xi_i$, which yields
the following:
\begin{equation}
    s^2V_{ij} = \delta^{kl}
		   (V_{kl}\xi_i\xi_j-V_{il}\xi_k\xi_j-V_{jl}\xi_k\xi_i
		+V_{ij}\xi_k\xi_l)
\end{equation}

\noindent or, equivalently
\begin{equation}
    s^2V_{ij} = V_{ij}+\xi_i\xi_jV -\xi_jV_{ik}\xi^k-\xi_iV_{jk}\xi^k\;,
\label{admlingeoeigen}
\end{equation}

\noindent where $V$ denotes the trace of $V_{ij}$ and indices are raised with
$\delta^{ij}$. The eigenvalues are thus labeled by $s^2$.  One can easily see
that this problem admits solutions with vanishing eigenvalue, as
follows. Assume $s=0$, which yields
\begin{equation}
    0 = V_{ij}+\xi_i\xi_jV -\xi_jV_{ik}\xi^k-\xi_iV_{jk}\xi^k
\label{admlingeoeigen1}
\end{equation}

\noindent Contracting with $\delta^{ij}$ we find $V-\xi^iV_{ij}\xi^j=0$.  Using
this information back into (\ref{admlingeoeigen1}) we have
\begin{equation}
    0 = V_{ij}+\xi_i\xi_j\xi^lV_{lm}\xi^m -\xi_jV_{ik}\xi^k-\xi_iV_{jk}\xi^k
\label{admlingeoeigen2}
\end{equation}

\noindent Now contracting with $\xi^j$ yields
\begin{equation}
0= \xi_i(V-\xi^lV_{lm}\xi^m),
\end{equation}

\noindent which are three identities. Therefore, three of the six components of
$V_{ij}$ are free, the remaining three being given by (\ref{admlingeoeigen2}). 
One can pick the three free components to be the three projections 
$V_{ij}\xi^j$, in which case, by (\ref{admlingeoeigen2}), the other three are
vanishing.  This observation leads directly to the fact that there are three
linearly independent eigenvectors with vanishing eigenvalue, the eigenvectors
being the six-dimensional unit vectors along the directions of the three
projections $V_{ij}\xi^j$. This is so for every arbitrary direction $\xi_i$. As
an illustration, representing the six-dimensional eigenvectors in the form
$\bm{u}\equiv (V_{xx},V_{xy},V_{xz},V_{yy},V_{yz},V_{zz})$, for
$\xi_i=\delta_i^x=(1,0,0)$ we have that $V_{ik}\xi^k=V_{xx},V_{xy},V_{xz}$ are
free, whereas, by (\ref{admlingeoeigen2}), the remaining components
$V_{yy},V_{yz}$ and $V_{zz}$ vanish. An arbitrary eigenvector in the degenerate
space of eigenvalue 0 has thus the form
\begin{eqnarray*}
{}^0\bm{u} &=& (V_{xx},V_{xy},V_{xz},0,0,0)\\
	   &=& V_{xx}(1,0,0,0,0,0) + V_{xy}(0,1,0,0,0,0)\\
&& 
		+V_{xz}(0,0,1,0,0,0),
\end{eqnarray*}

\noindent where $V_{xx},V_{xy}$ and $V_{xz}$ are arbitrary real numbers. There
are thus three zero-speed eigenvectors
associated with the $x-$direction which can be chosen as
\begin{eqnarray*}
	{}^0\bm{u_1}&=&(1,0,0,0,0,0),\\
	{}^0\bm{u_2}&=&(0,1,0,0,0,0),\\
	{}^0\bm{u_3}&=&(0,0,1,0,0,0),
\end{eqnarray*}

\noindent which are manifestly linearly independent. In accordance with the
criterion of Section~\ref{sec:1}, this is enough to demonstrate that the
evolution equations (\ref{admlingeo}) are not strongly hyperbolic. 
Nonetheless, for completeness, one can calculate the remainder of the
eigenvalues and eigenvectors in a similar manner (or one could, of course,
proceed by any standard methods of linear algebra to calculate all the
eigenvectors and eigenvalues at the same time). The reader can verify that
the only other eigenvalue is $s^2=1$, for which Eq.~(\ref{admlingeoeigen}) reads
\begin{equation}
	0= \xi_i\xi_jV -\xi_jV_{ik}\xi^k-\xi_iV_{jk}\xi^k\;.
\end{equation}

\noindent These six equations are all satisfied if and only if
\begin{equation}\label{Vlong}
	V_{ik}\xi^k = \xi_i\frac{V}{2}.
\end{equation}

\noindent This in turn implies that the three components of $V_{ij}$ other than
the projections $V_{ik}\xi^k$ are arbitrary. Denoting them by $V^\perp_{ij}$
they are given by
\begin{equation}\label{Vperp}
V^\perp_{ij} \equiv
V_{ij}-\xi_iV_{jk}\xi^k-\xi_jV_{ik}\xi^k+\xi_i\xi_i\xi^kV_{kl}\xi^l
\end{equation}

\noindent and are such that $V^\perp_{ij}\xi^j=0$ by construction (which means
that there are only three independent components in $V^\perp_{ij}$).  This
leads directly to the conclusion that there are three linearly independent
eigenvectors with light speed, labelled by the three components of $V_{ij}$
\textit{other than} $V_{ik}\xi^k$, and they are, clearly, linearly independent
from the zero-speed eigenvectors (labelled by $V_{ik}\xi^k$). For instance, in
the case that $\xi_i=(1,0,0)$ as above, by (\ref{Vperp}) the three arbitrary
components of $V_{ij}$ are $V_{yy}, V_{yz}$ and $V_{zz}$, which, if combined
with (\ref{Vlong}), leads to the fact that an arbitrary eigenvector in the
three-dimensional degenerate space of eigenvalue 1 is of the form
\begin{eqnarray*}
{}^1\bm{u} &=& (V_{yy}+V_{zz},0,0,V_{yy},V_{yz},V_{zz})\\
	   &=& V_{yy}(1,0,0,1,0,0) + V_{yz}(0,0,0,0,1,0)\\
&& 
		+V_{zz}(1,0,0,0,0,1).
\end{eqnarray*}

\noindent where $V_{yy}, V_{yz}$ and $V_{zz}$ are completely arbitrary real
numbers. This shows that three unit eigenvectors can be chosen as
\begin{eqnarray*}
	{}^1\bm{u_1}&=&\frac{1}{\sqrt{2}}(1,0,0,1,0,0),\\
	{}^1\bm{u_2}&=&(0,0,0,0,1,0),\\
	{}^1\bm{u_3}&=&\frac{1}{\sqrt{2}}(1,0,0,0,0,1),
\end{eqnarray*}

\noindent which are manifestly linearly independent of each other and of 
${}^0\bm{u_i}$.

In summary,  the principal symbol $P_0(\xi)$ of (\ref{admlingeo}) admits a
complete set of eigenvectors and has real but not strictly positive
eigenvalues.  The linearized second-order ADM equations with constant unit
lapse are, thus, weakly hyperbolic. Consequently, the original nonlinear
second-order ADM equations (\ref{admsecond}) are prone to catastrophic
exponential growth, as is their linearization around any background but flat. 

Suppose now that the lapse function is equal to the square root of the
determinant of the three-metric.  In the linearization, this means that
$2\epsilon,_{ij} = \delta^{kl}h_{kl,ij}$. With this choice of lapse, 
Eq.~(\ref{admlin}) reads
\begin{equation}\label{admlinharm}
    \ddot{h}_{ij} = \delta^{kl}
		   (2h_{kl,ij}-h_{il,kj}-h_{jl,ki}+h_{ij,kl}) 
\end{equation} 

\noindent The associated eigenvalue problem is
\begin{equation}
    s^2V_{ij} = V_{ij}+2\xi_i\xi_jV -\xi_jV_{ik}\xi^k-\xi_iV_{jk}\xi^k
\label{admlinharmeigen}
\end{equation}

\noindent One can see that this eigenvalue problem admits two linearly
independent solutions for $s=0$.  To see this quickly (without necessarily
using  standard algebraic methods to solve the problem completely), set $s=0$
and contract with $\delta^{ij}$, which yields $V = (2/3) \xi^iV_{ij}\xi^j$.
Substituting this back and contracting this time with $\xi^j$ one has
\begin{equation}\label{xivxi}
   0 = \frac13\xi_i\xi^lV_{lm}\xi^m 
\end{equation}

\noindent Thus four out of the six equations are solved by setting
$\xi^iV_{ij}\xi^j=0$ and consequently also $V=0$. This implies that two
components of $V_{ij}$ are free, which can be taken as the two $V_{ij}\xi^j$
other than  $\xi^iV_{ij}\xi^j$.  The remaining two equations in the set fix the
remaining two components of $V_{ij}$ in terms of these:
\begin{equation}\label{Vij}
    0= V_{ij} -\xi_jV_{ik}\xi^k-\xi_iV_{jk}\xi^k
\end{equation}

\noindent As in any eigenvalue problem, the fact that two components of
$V_{ij}$ are left arbitrary by  (\ref{admlinharmeigen}) with $s=0$ leads
directly to the conclusion that there are two linearly independent
eigenvectors (which can be calculated in any way the reader finds appealing)
associated with zero speed and any direction $\xi_i$. For instance, if
$\xi_i=(1,0,0)$, then $V_{xy}$ and $V_{xz}$ are free, but $V_{xx}=0$ by
(\ref{xivxi}), and  $V_{yy}, V_{yz}$ and $V_{zz}$ also vanish by virtue of
(\ref{Vij}). The generic eigenvector in the degenerate space of $s=0$ is
thus
\begin{eqnarray*}
{}^0\bm{u} &=& (0,V_{xy},V_{xz},0,0,0)\\
	&=& V_{xy}(0,1,0,0,0,0)+V_{xz}(0,0,1,0,0,0),
\end{eqnarray*}

\noindent where $V_{xy}$ and $V_{xz}$ are arbitrary real numbers, so there are
only two linearly independent eigenvectors and they can be chosen as
\begin{eqnarray*}
	{}^0\bm{u_1}&=&(0,1,0,0,0,0),\\
	{}^0\bm{u_2}&=&(0,0,1,0,0,0).
\end{eqnarray*}

\noindent This is
enough to conclude that the linearized second-order ADM equations with
``harmonic'' lapse, namely Eq.~(\ref{admlinharm}), are not strongly hyperbolic.

For completeness, one can calculate the remaining eigenvalues and
eigenvectors.  The reader can verify that the only eigenvalue other than zero
is $s^2=1$, which has three eigenvectors associated with it, namely the
components of $V_{ij}$ other than $V_{ik}\xi^k$.  In keeping up with our
illustration but skipping over the procedure, which is entirely similar to the
one used three times in the preceeding, for $\xi_i=(1,0,0)$ the three
eigenvectors associated with light speed are (i.e., can be chosen as)
\begin{eqnarray*}
	{}^1\bm{u_1}&=&\frac{1}{\sqrt{3}}(1,1,1,0,0,0),\\
	{}^1\bm{u_2}&=&(0,0,0,0,1,0),\\
	{}^1\bm{u_3}&=&\frac{1}{\sqrt{2}}(0,0,0,1,0,-1),
\end{eqnarray*}

\noindent which are manifestly linearly independent of each other and of 
${}^0\bm{u_1}, {}^0\bm{u_2}$.   Because the eigenvalues are real and
non-negative and there is not a complete set of eigenvectors,
Eqs.~(\ref{admlinharm}) are weakly hyperbolic. One can infer that the
second-order nonlinear ADM equations (\ref{admsecond}) are prone to
``catastrophic exponential growth'' even in the case of a lapse function
proportional to the determinant of the three-metric, as is their linearization
around any nonflat background. 

This result does not contradict~\cite{ortiz}.  In \cite{ortiz}, the authors do
consider the linearized second-order ADM equations and conclude that in the
case of a lapse function that is proportional to the determinant of the three
metric the problem is strongly hyperbolic. However, they impose, additionally,
the linearized constraints on the evolution equations, which, as a consequence,
do not have the form (\ref{admlinharm}).  Thus the problem that  Kreiss and
Ortiz found to be well posed is equivalent to a constrained evolution problem
for the ADM equations in second-order with ``harmonic'' lapse, where the
solutions are constrained at every time slice. In contrast, we have
demonstrated here that the corresponding unconstrained evolution problem is not
well posed.

\section{The BSSN equations in second order form\label{sec:3}}

In the case of vanishing shift vector, the formulation of the
Einstein equations referred to as BSSN~\cite{BSSN} consists of the
following evolution equations
\begin{subequations}\label{bssn}
\begin{eqnarray}
\dot{\tilde{\gamma}}_{ij} &=& -2\alpha \tilde{A}_{ij} \label{bssndotg}
\\
\dot{\phi} &=& -\frac{\alpha}{6} K \label{bssndotphi}
\\
\dot{K} &=& -\gamma^{ij}D_iD_j\alpha + \alpha(\tilde{A}_{ij}\tilde{A}^{ij}
    +\frac13K^2)			\label{bssndotK}
\\
\dot{\tilde{\Gamma}}^i &=&
    2\alpha\Big(\tilde{\Gamma}^i{}_{jk}\tilde{A}^{kj}
+\frac23\tilde{\gamma}^{ij}K,_j
     + 6 \tilde{A}^{ij}\phi,_j\Big)
\nonumber\\ &&
-2\tilde{A}^{ij}\alpha,_j		\label{dotGamma}
\\
\dot{\tilde{A}}_{ij} &=&\alpha e^{-4\phi} \bigg(-\frac12 \tilde{\gamma}^{lm}\tilde{\gamma}_{ij,lm}
    +\tilde{\gamma}_{k(i}\tilde{\Gamma}^k,_{j)}
    + \tilde{\Gamma}^k\tilde{\Gamma}_{(ij)k} \nonumber\\
&&
    +2\tilde{\Gamma}^{kl}{}_{(i}\tilde{\Gamma}_{i)kl}
    +\tilde{\Gamma}^{kl}{}_i\tilde{\Gamma}_{klj}
    -2\tilde{D}_i\tilde{D}_j\phi
    +4\tilde{D}_i\phi\tilde{D}_j\phi\nonumber\\
&&
    -\frac13\tilde{\gamma}_{ij}\big(\tilde{\Gamma}^k,_k
    +\tilde{\Gamma}^{kli}(2\tilde{\Gamma}_{ikl}
                  +\tilde{\Gamma}_{kli})\big)\nonumber\\
&&
    +\frac23\tilde{\gamma}_{ij}(\tilde{D}^l\tilde{D}_l\phi
    -2\tilde{D}^l\phi\tilde{D}_l\phi)
-{\frac{(D_iD_j\alpha)}{\alpha}}^{TF}\bigg)\nonumber\\
&&
            +K\tilde{A}_{ij}
-2\tilde{A}_{il}\tilde{A}^l_j \label{bssndotA}
\end{eqnarray}
\end{subequations}

\noindent for the 15 variables
\begin{subequations}
\begin{eqnarray}
\phi&\equiv& \frac{1}{12}\ln (\det \gamma_{ij})\\
\tilde{\gamma}_{ij} &\equiv& e^{-4\phi}\gamma_{ij}\\
K &\equiv& \gamma^{ij}K_{ij}\\
\tilde{A}_{ij} &\equiv& e^{-4\phi}\left(K_{ij}
        -\frac13\gamma_{ij}K\right)\\
\tilde{\Gamma}^i &\equiv& -\tilde{\gamma}^{ij},_j
\end{eqnarray}
\end{subequations}

\noindent As in the ADM case, the initial data for these evolution equations
must be chosen to satisfy the constraints (\ref{admconst}). Other than as
applied to the initial data, we will disregard the constraints.  

We obtain a six-dimensional second order system for the metric variables by
substituting $A_{ij}$ in terms of $\dot{\tilde{\gamma}}_{ij}$ back into
(\ref{bssndotA}), and $K$ in terms of $\dot{\phi}$ into (\ref{bssndotK}).
Additionally, we substitute $\tilde{\Gamma}^i$   back
in terms of $\tilde{\gamma}^{ij},_j$ into (\ref{bssndotA}). 

The linearization around flat space $\gamma_{ij} = \delta_{ij} + h_{ij}$ implies
that $\tilde{\gamma}_{ij} = \delta_{ij} + \tilde{h}_{ij}$ and $\phi = h/12$,
where $\tilde{h}_{ij} = h_{ij} - (1/2)\delta_{ij} h$ and $h=\delta^{ij}h_{ij}$.  
After linearization around flat space, the result is the following second-order
system 
\begin{subequations}\label{bssnlin}
\begin{eqnarray}
\ddot{\phi} &=& \frac16 \delta^{ij}\epsilon,_{ij}\\
\ddot{\tilde{h}}_{ij} &=& \delta^{kl}\left(\tilde{h}_{ij,kl}
				   -\tilde{h}_{il,kj}
				   -\tilde{h}_{jl,ki}
	+\frac23 \delta_{ij}\delta^{rs}\tilde{h}_{ls,kr} \right)\nonumber\\
&& + 4\Big(\phi,_{ij} -\frac13\delta_{ij}\delta^{kl}\phi,_{kl}\Big)
+ 2\Big(\epsilon,_{ij}
-\frac13\delta_{ij}\delta^{kl}\epsilon,_{kl}\Big).\nonumber\\
&&
\end{eqnarray}
\end{subequations}

Consider first the case of constant lapse equal to 1. The equations reduce to
\begin{subequations}\label{bssnlingeo}
\begin{eqnarray}
\ddot{\phi} &=& 0\\
\ddot{\tilde{h}}_{ij} &=& \delta^{kl}\left(\tilde{h}_{ij,kl}
				   -\tilde{h}_{il,kj}
				   -\tilde{h}_{jl,ki}
	+\frac23 \delta_{ij}\delta^{rs}\tilde{h}_{ls,kr} \right)\nonumber\\
&& + 4\Big(\phi,_{ij} -\frac13\delta_{ij}\delta^{kl}\phi,_{kl}\Big),
\end{eqnarray}
\end{subequations}

\noindent and the corresponding eigenvalue problem, with 
$\tilde{h}_{ij}= \tilde{V}_{ij}\exp i(\xi_kx^k -st)$ and 
$\phi= \psi\exp i(\xi_kx^k -st)$, is 
\begin{subequations}
\begin{eqnarray}
s^2\psi &=& 0 \label{psi}\\
s^2\tilde{V}_{ij} &=& \tilde{V}_{ij}
		      -\xi_i\tilde{V}_{jl}\xi^l
		      -\xi_j\tilde{V}_{il}\xi^l
	+\frac23 \delta_{ij}\xi^k\tilde{V}_{kl}\xi^l\nonumber\\
&& + 4\psi\xi_i\xi_j -\frac43\delta_{ij}\psi\;.\label{tildeV}
\end{eqnarray}
\end{subequations}

\noindent This system has a vanishing eigenvalue $s^2=0$ with three linearly
independent associated eigenvectors, labelled by the three components
$\tilde{V}_{ij}\xi^j$. In order to see this, set $s^2=0$ in (\ref{tildeV}) and
contract with $\xi^i$. This yields $\xi_j(\psi - (1/8)
\xi^k\tilde{V}_{kl}\xi^l)=0$. So three equations are satisfied by the choice
$\psi = (1/8) \xi^k\tilde{V}_{kl}\xi^l$. Additionally, since $s^2=0$, then also
(\ref{psi}) is identically satisfied.  Four out of the six equations are thus
satisfied with this one choice, which means that three of the fields are free.
This is enough to conclude that the second-order BSSN equations with unit
lapse, namely Eq.~(\ref{bssnlingeo}), are not strongly hyperbolic.  We can
calculate the remaining eigenvalues and eigenvectors.  The reader can verify
that there are two other eigenvalues.  One is $s^2=1$, with two associated
eigenvectors, which are the components of $\tilde{V}_{ij}$ other than
$\tilde{V}_{il}\xi^l$, with $\tilde{V}_{il}\xi^l=0$ and $\psi=0$. The other
eigenvalue is negative $s^2=-1/3$, and has one eigenvector associated with it,
which is given by $\psi=0$, $\tilde{V}_{il}\xi^l=0$ except for 
$\xi^i\tilde{V}_{il}\xi^l$ which is free, and $\tilde{V}_{ij} =
(3/2)(\xi_i\xi_j-(1/3)\delta_{ij})\xi^k\tilde{V}_{kl}\xi^l$. Thus there is a
complete set of eigenvectors and all eigenvalues are real, but are not
non-negative, and the equations are not even weakly hyperbolic. This means that the full nonlinear second-order
BSSN equations with unit lapse have the potential for ``catastrophic growth'',
as does their linearization around any background other than flat. 

Things change very little if one considers now the case of $\alpha= \sqrt{\det
\gamma_{ij}}$, or $\epsilon,_{ij} = 6\phi,_{ij}$.  Equations (\ref{bssnlin})
reduce to 
\begin{subequations}\label{bssnlinharm}
\begin{eqnarray}
\ddot{\phi} &=& \delta^{ij}\phi,_{ij}\\
\ddot{\tilde{h}}_{ij} &=& \delta^{kl}\left(\tilde{h}_{ij,kl}
				   -\tilde{h}_{il,kj}
				   -\tilde{h}_{jl,ki}
	+\frac23 \delta_{ij}\delta^{rs}\tilde{h}_{ls,kr} \right)\nonumber\\
&& + 16\Big(\phi,_{ij} -\frac13\delta_{ij}\delta^{kl}\phi,_{kl}\Big).
\end{eqnarray}
\end{subequations}

\noindent This system has two eigenvectors with $s^2=0$, which are given by
$\xi^i\tilde{V}_{il}\xi^l=0$, $\psi=0$ and $\tilde{V}_{ij}=
\xi_i\tilde{V}_{jl}\xi^l + \xi_j\tilde{V}_{il}\xi^l$ with arbirary values for
the two components $\tilde{V}_{il}\xi^l$ other than 
$\xi^i\tilde{V}_{il}\xi^l$.  There are two other eigenvalues different from
zero. One is $s^2=1$, with three associated eigenvectors which have free values
of $\psi$ and of the components of $\tilde{V}_{ij}$ other than
$\tilde{V}_{ij}\xi^j = 8\xi_i\psi$. The other one is negative $s^2=-1$, and has
one eigenvector associated with it which has $\psi=0$ and $\tilde{V}_{ij} =
(3/2)(\xi_i\xi_j-(1/3)\delta_{ij})\xi^k\tilde{V}_{kl}\xi^l$ with 
$\xi^k\tilde{V}_{kl}\xi^l$ arbitrary.  So (\ref{bssnlinharm}) has a complete
set of eigenvectors and its eigenvalues are real but not non-negative, and thus
it is not even weakly hyperbolic. The consequences to the full nonlinear BSSN
equations in second-order form and with ``harmonic'' lapse are the same as in
the case of unit lapse. 


\section{The generalized EC equations in second order form\label{sec:4}}

The generalized Einstein-Christoffel (EC) formulation is first introduced in
\cite{KST}, where the method of derivation from the ADM equations is described
without the details of the resulting equations themselves. The principal part
of the evolution equations of the system appear explicitly
in \cite{calabrese3d} as follows
\begin{subequations}
\begin{eqnarray}
\dot{\gamma}_{ij} &=& -2\alpha K_{ij}\\
\dot{K}_{ij} &=& -\alpha\gamma^{kl}\partial_lf_{kij} + \ldots \label{kdot}\\
\dot{f}_{kij} &=& -\alpha\partial_k K_{ij} + \dots \label{fdot}
\end{eqnarray}
\end{subequations}

\noindent where $f_{kij}$ constitute a set of 18
first-order variables defined by the following relationship with the first
derivatives of the three-metric~\cite{calabrese3d}:
\begin{equation}\label{fkij}
\gamma_{ij,k} \equiv 2 f_{kij} 
		+    \eta \gamma_{k(i}\left(f_{j)s}{}^s - f^s{}_{j)s}\right)
		+ \frac{\eta-4}{4}\gamma_{ij}(f_{ks}{}^s - f^s{}_{ks}).
\end{equation}

\noindent Here $\eta$ is a free parameter.  The generalized EC family requires
the lapse function to be densitized, that is: to be proportional to the square
root of the determinant of the three-metric. The original (standard) EC
formulation obtained by Anderson and York~\cite{fixing} corresponds to the
choice of $\eta=4$. 

We are interested in the second-order system of equations for $\gamma_{ij}$
that is implied by this 30-dimensional first-order problem. We start by
inverting (\ref{fkij}) in order to have an expression for $f_{kij}$ that we can
use to substitute in terms of $\gamma_{ij}$.  The inversion yields
\begin{eqnarray}\label{ftogamma}
2f_{kij}  &=& \gamma_{ij,k}
	     + \gamma_{ik}( \gamma^{ls}\gamma_{ls,j}
			   -\gamma^{ls}\gamma_{lj,s})\nonumber\\
&&	     + \gamma_{jk}( \gamma^{ls}\gamma_{ls,i}
			   -\gamma^{ls}\gamma_{li,s})\nonumber\\
&&	     +\frac{\eta-4}{2\eta}\gamma_{ij}( \gamma^{ls}\gamma_{ls,k}
			   		      -\gamma^{ls}\gamma_{lk,s}).
\end{eqnarray}
	 
If one uses (\ref{ftogamma}) in its left-hand side,  Eq.~(\ref{fdot}) reduces to
a linear combination of the components of the momentum constraint ${\cal C}_i$. 
Therefore, this equation is redundant for the problem that we are interested in.
However, eliminating $f_{kij}$ from the right-hand side of (\ref{kdot}) by means
of (\ref{ftogamma}) yields:
\begin{equation}\label{premodEC}
\dot{K}_{ij} = \alpha \left(R_{ij} - 2 K_{il}K^l{}_j
            +K K_{ij}\right) -D_iD_j\alpha +
\frac{\eta-4}{2\eta}\alpha\gamma_{ij} {\cal C}.
\end{equation}  

\noindent One can see that, at $\eta=4$, the evolution equation for the
extrinsic curvature in the standard EC formulation is an
exact transcription of the ADM evolution equation without mixing of the
constraints, whereas for any other value of the parameter $\eta$, the
Hamiltonian constraint is involved. This means that the second-order version of
the standard EC is indistinguishable from the second-order version of the ADM
equations, dealt with in Section~\ref{sec:2}. Additionally, one can also show
that for $\eta=12/7$, the second-order version of the generalized EC
formulation is indistinguishable from the second-order version of the BSSN
formulation, which is dealt with in Section~\ref{sec:3}. Yet, for $\eta\neq 4,
12/7$, the second-order version of the generalized EC is a genuinely different
problem which requires separate study. 

The second-order problem that we seek is found by substituting $K_{ij}$ in terms
of $\dot{\gamma}_{ij}$ in the left-hand side of (\ref{premodEC}) and writing
explicitly the right-hand side in terms of $\gamma_{ij,kl}$:
\begin{eqnarray}
\alpha^{-2}\ddot{\gamma}_{ij} &=& 
		     	  \gamma^{kl}\gamma_{ij,kl}
			- \gamma^{kl}\gamma_{il,kj}
			- \gamma^{kl}\gamma_{kj,il}
			+2\gamma^{kl}\gamma_{kl,ij}\nonumber\\
&&
		+\frac{\eta-4}{2\eta}\gamma_{ij}(
			  \gamma^{kl}\gamma^{ms}\gamma_{ms,kl}
			- \gamma^{kl}\gamma^{ms}\gamma_{ml,ks})\nonumber\\
&&		+\dots
\end{eqnarray}

In the linearization around flat space ($\gamma_{ij} = \delta_{ij} + h_{ij}$)
we have
\begin{eqnarray}
\ddot{h}_{ij} &=& 
		     	  \delta^{kl}h_{ij,kl}
			- \delta^{kl}h_{il,kj}
			- \delta^{kl}h_{kj,il}
			+2\delta^{kl}h_{kl,ij}\nonumber\\
&&
		+\frac{\eta-4}{2\eta}\delta_{ij}(
			  \delta^{kl}\delta^{ms}h_{ms,kl}
			- \delta^{kl}\delta^{ms}h_{ml,ks}),
\end{eqnarray}

\noindent which has the following eigenvalue problem in terms of $h_{ij} =
V_{ij} \exp i(\xi_kx^k -st)$: 
\begin{eqnarray}
s^2 V_{ij} &=&  V_{ij}
		- V_{ik}\xi^k\xi_j
		- V_{jk}\xi^k\xi_i
		+2 V\xi_i\xi_j\nonumber\\
&&
		+\frac{\eta-4}{2\eta}\delta_{ij}(
			  V
			- \xi^kV_{kl}\xi^l).
\end{eqnarray}

\noindent This problem has the following eigenvalues. First we have $s^2=1$
with three linearly independent eigenvectors which are the three components of
$V_{ij}$ other than the projections $V_{ij}\xi^j (=\xi_iV)$.  Then we have
$s^2=2(\eta-2)/\eta$ with one eigenvector given by $V_{ij} = 2\eta V
(2\xi_i\xi_j+\delta_{ij}(\eta-4)/2\eta)/(7\eta-12)$ with arbitrary $V$, which
is linearly independent of the other three except at $\eta=4$.  Finally we have
$s^2=0$ with two eigenvectors given by $V_{ij} =
\xi_iV_{jk}\xi^k+\xi_jV_{ik}\xi^k$ with $V = \xi^kV_{kl}\xi^l=0$.  

So there is a complete set of eigenvectors for all values of $\eta\neq 4,
12/7$. However, there are two eigenvectors with $s^2=0$ for all $\eta\neq 2, 4,
12/7$.  Furthermore, if $\eta=2$ then $s^2=0$ has multiplicity 3, but if $\eta
< 2$ then there is a negative eigenvalue.  Collectively, this means that for no
values of $\eta$ is the second-order problem implied by the generalized EC
formulation strongly hyperbolic. The second-order problem is weakly hyperbolic
for $\eta\ge 2$, which includes the ADM case, and this is consistent with
Section~\ref{sec:2}. Yet the second-order problem is not even weakly hyperbolic
in the range $\eta < 2$ considered in ~\cite{calabrese3d}, which includes the
BSSN case, and this is consistent with Section~\ref{sec:3}.

\section{Concluding remarks\label{sec:5}}

Summarizing, we have found that the second-order equations for the three-metric
that are implied by a large number of first-order reductions of the 3+1
Einstein equations whose evolution equations differ by the addition of
different multiples of the Hamiltonian constraint are potentially ill-posed
irrespective of whether the lapse is constant or densitized. We say
\textit{potentially\/} only because the linearized argument that we use does in
on way rule out the possibility that the lower-order terms that are present in
the nonlinear equations may yet (fortituously) result in a well-posed nonlinear
problem. This should raise a flag for numerical efforts seeking formulations
for general relativity with the smallest possible number of variables.  

On the other hand, studying the second-order equations implied by formulations
with such disparity of properties as ADM, BSSN and generalized EC gives us yet
another perspective for what it is about such formulations that characterizes
their properties. 

In the case of the ADM equations and their associated second-order version, we
have verified a fact that has been suspected for a long time, starting with
\cite{helmut96}, where the second-order version itself is found unfit for
standard hyperbolicity studies.  Subsequently, many authors have concluded that
the ADM equations themselves are ill posed on the basis of the existence of
full first-order reductions not involving constraint mixing in the evolution
equations of the first-order variables. This intuitive line of reasoning,
whether rigorously justified or not, naturally leads to the conclusion that the
second-order version should also be ill-posed. Our results of
Section~\ref{sec:2} provide a rigorous basis for this intuitive inference.

In the case of the BSSN equations the implications are less straightforward. In
essence, the BSSN equations themselves constitute a partial first-order
reduction of the ADM equations with constraint mixing in two senses. First,
there is the use of a multiple of the Hamiltonian constraint in the evolution
equation for the extrinsic curvature. Secondly, there is a use of the momentum
constraint in the evolution equations for the three new first-order variables
$\tilde{\Gamma}^i$.  According to \cite{NOR}, the latter play a critical role
in the well-posed properties of the BSSN equations in a pseudospectral sense,
whereas the former is irrelevant to well-posedness in such a sense.  This
accounts for the difference in the well-posedness of the first-order reduction
of BSSN as compared to ADM (but of course, the comparison is not completely
fair because BSSN is already a partial reduction).  Yet, the constraint mixing
in the evolution equations of the new first-order variables plays no role
whatsoever in the well-posedness of the second-order problem for the
three-metric components because it simply leads to a redundancy, as explained
in the main body of this article.  Therefore, it is disappointing but hardly
surprising that the second-order version of the BSSN equations does not turn
out to be well posed.   In fact, this verifies the idea put forward in
\cite{NOR} that the constraint mixing in the evolution of the first-order
variables is crucial to the well-posedness of a first-order reduction of the
Einstein equations.  

This idea is most strongly upheld by the results in the case of the generalized
EC formulation and its second-order version. Here is a family of first-order
reductions all of which are well posed for any multiple of the Hamiltonian
constraint added to the evolution equation for the extrinsic curvature.  But
they all have momentum-constraint mixing in the evolution equations for the new
first-order variables $f_{kij}$. Yet none of the second-order versions of these
formulations turns out to be well posed. What the second-order versions are
missing is precisely the evolution equations for the first-order variables
(which become redundant in the second-order version).  
For several reasons including the present ones, the constraints seem to be
surfacing as key players in the initial value problem of the Einstein
equations, contrary to their long-standing reputation as choosers of physical
initial data but otherwise ignorable.   

Additionally, reflecting on the fact that the standard second-order version of
the Einstein equations with harmonic slicing where all the components of the
metric evolve according to wave equations is indeed well posed, one is led to
conclude that the shift vector must play a crucial role in the well-posedness
of second-order versions of the 3+1 Einstein equations, in agreement with the
intuition of the authors of \cite{bonapseudo} where a dynamical shift choice is
used to analyse some non-standard hyperbolicity properties of the BSSN
equations. This remains an open problem.

\begin{acknowledgments}
This work was supported by the NSF under grant No.
PHY-0244752 to Duquesne University.
\end{acknowledgments}


\end{document}